\documentclass[12pt]{article}%

\begin{document}
\title{Families of Canonical Transformations by Hamilton-Jacobi-Poincar\'e equation. Application to  Rotational and Orbital Motion}
\author{S. Ferrer\thanks{Departamento de Matem\'atica Aplicada, Universidad de Murcia,
30071 Espinardo, Spain} \, and \,  M. Lara\thanks{Real Observatorio de la Armada, 11110 San Fernando, Spain}
}

\maketitle

\begin{abstract}

The Hamilton-Jacobi equation in the sense of Poincar\'e, i.e. formulated in the extended phase space and including regularization, is revisited building canonical transformations with the purpose of Hamiltonian reduction. We illustrate our approach dealing with orbital and attitude dynamics. Based on the use of Whittaker and Andoyer symplectic charts, for which all but one coordinates are cyclic in the Hamilton-Jacobi equation, we provide whole families of canonical transformations, among which one recognizes the familiar ones used in orbital and attitude dynamics. In addition, new canonical transformations are demonstrated.

\end{abstract}


\section{Introduction}

Transformation of variables, including or not the change of time used as evolution parameter, is a basic tool in almost any problem of physics. A recent illustration of the usefulness of this technique, which original application goes back to Laplace and Euler, may be found in \cite{Grandati}. Changes of variables belong to Geometry and, therefore, do not depend on the physical system to which they may be applied. However, transformations are commonly derived for specific purposes and, hence, they are tightly related to the particular problem they are applied to. 
\par

A renewed interest on transformation of variables tailored to specific problems has recently emerged in three areas. On one hand there is a continuous search for more efficient numerical integrators demanded in many applied fields, ranging from research groups endowed with powerful computing facilities, to research involved in very long periods of time (see \cite{BlanesBudd,Fukushima2008a} and references therein). On the other hand the Hamilton-Jacobi (H-J) equation is receiving new attention in extended phase space, connecting this technique with the variational principles of Lagrangian and Hamiltonian Mechanics (see \cite{Struckmeier} and references therein). Finally a third area, which has its roots in Poincar\'e's \cite{Poincare} \textsl{fundamental problem of dynamics}, deals with Hamiltonian systems of the form $\mathcal{H}=\mathcal{H}_0+\mathcal{H}_1$ where $\mathcal{H}_0$ defines an integrable system, and $\mathcal{H}_1$ is considered a perturbation. The H-J equation is built from $\mathcal{H}_0$ by searching for a generating function such that the new canonical variables satisfy some requirements of perturbation theories (see, for instance, \cite{Yanguas,Fukushima2008b}). The content of this paper focuses in this last area. Finally note that the H-J approach has been used by many authors, along the history of this equation, looking for variables leading to separability. We do not consider that problem as such, although it might be included.

There is a long tradition in Classical and Celestial Mechanics ---nowadays a common ground in many other fields--- of using both canonical transformations and changes of independent varible dubbed as regularization. The H-J equation is the usual approach, in which the canonical transformation is obtained from either a \emph{principal} or a \emph{characteristic} function \cite{Goldstein2003}.
\par

The characteristic function approach does not rely on the vanishing of the new Hamiltonian. It only requires a non-zero transformed Hamiltonian that is cyclic in all variables but depends on one or several momenta. The adequate selection of the new Hamiltonian will make the transformation successful for specific purposes. Thus, for instance, if one wants the transformed Hamiltonian to retain the topology of the original problem, probably several momenta must be retained in the new Hamiltonian. On the contrary, useful transformations for perturbation theory may look for transformed Hamiltonians that depend only on one momenta, thus simplifying the subsequent application of perturbation methods (which, in turn, have limited application to specific solutions as, for instance, quasi-periodic motion).
\par

Sentences like ``after several trials'', which are frequent in the literature, might produce the impression that succeeding in the choice of the reduced Hamiltonian is a matter of educated guess or \textit{alquimia}. However, the new Hamiltonian does not need to be specified since the beginning and, quite on the contrary, it may be handled formally throughout the process of computing the transformation \cite{SussmanWisdom}. Therefore, Hamiltonian problems solved by the H-J equation give rise to whole families of canonical transformations from which the selection of a particular one requires only the materialization of the new Hamiltonian\footnote{To make more clear our point from \cite{SussmanWisdom} (p.~412), where $E$ denotes the Hamiltonian in the new variables, we quote ``We are still free to choose the functional form of $E$. A convenient (and conventional) choice is $E(p_0,p_1,p_2)=-m\mu^2/(2p_0^2)$. With this choice the momentum $p_0$ has dimensions of angular momentum, and the conjugate coordinate is an angle.'' As we will show this is not the only way of reasoning. Indeed, we arrive to the same choice when we impose the function called {\sl Kepler equation} as one of the expressions defining the transformation. In other words, when the {\sl mean anomaly} is chosen as one of the new coordinates. As Delaunay showed, in doing so he avoided the presence of mixed secular terms in this  theory of the Moon. For a recent study on these variables from a geometric point of view the reader should consult \cite{ChangMarsden2003}}.
\par

We propose to  obtain  canonical transformations from a general form of a \mbox{H-J} equation that is formulated in the extended phase space. In principle, the only requirement imposed to the Hamiltonian function (\ref{ham1}) is that $\mathcal{H}_0$ has to be integrable.  More precisely, the two families studied are presented using symplectic charts where the unperturbed part defines a 1-DOF. 
In other words, we only deal with canonical transformations derived from integrable Hamiltonians that are cyclic in all but one variable, and perform two specific applications. 
In attitude dynamics, we find a family of transformations that affords for complete reduction of the Euler-Poinsot problem. In this case, we check that Sadov's \cite{Sadov,SadovRuso}, Kinoshita's \cite{Kinoshita72}, Deprit's \cite{DepritElipe1993}, and the recent Fukushima's \cite{Fukushima2008b} reduced Hamiltonians are just particular cases of our family. Notably, our approach provides explicit trans\-for\-ma\-tions from Sadov's and Kinoshita's elements to Andoyer variables \cite{AndoyerBook}, thus overcoming the major objection to both sets of elements of being constructed through implicit relations. Besides, we show that other selections of the new Hamiltonian provide new canonical transformations that may help in simplifying perturbation algorithms in attitude dynamics.

In orbital dynamics we obtain three different families of \textit{isochronal} canonical transformations ---derived from H\'enon's isochronal central potential \cite{Henon,BoccalettiPucacco}--- each one linked with a different form of the regularizing function. From our families we recover a variety of canonical transformations in the literature, ranging from historic ones like Delaunay \cite{Delaunay}, Levi-Civita \cite{LeviCivita1,LeviCivita2}, or Hill \cite{Hill}, to the recent transformation of Yanguas \cite{Yanguas}. In addition, we show that the set of Delaunay elements yields the simplest transformation of the isochronal family that reduces the Hamiltonian of the two body problem to one momentum and, besides, preserves the original form of the Kepler equation.
\par

The two charts used enjoy similar characteristics. Both express the unperturbed part of the Hamiltonian families as genuine 1-DOF systems due to the central and axial symmetries that both possess. They belong to the category of superintegrable systems, but not maximally superintegrable (see Fass\`o \cite{Fasso} and Ortega and Ratiu \cite{OrtegaRatiu}); only for particular choices of the parameters both systems fall into this last category. Moreover, from the computational point of view, these charts introduce great simplification in intermediary expressions which, in turn, lead to a deep insight about each family. Many authors still rely on Eulerian variables when dealing with rotational dynamics  (see for instance Marsden and Ratiu \cite{MarsdenRatiu}) , and spherical variables for orbital type motions (see Goldstein {\sl et al.} \cite{Goldstein2003}, Jos\'e and Saletan \cite{JoseSaletan} and Sussman and Wisdom \cite{SussmanWisdom}).\par
In spite of the fact that the two families considered are independent and consequently, they are studied in a totally selfcontained manner, we make both part of the paper because this allows to obtain a better insight in the way we approach the H-J formalism.
A final comment is due. Although both families have parameters, we have made the analysis for the generic case only. With respect to the orbital case, as an illustration, we have added details when we restrict to  Keplerian systems, showing the connection of the new variables with the {\sl anomalies}. We would like to make clear that we are not doing a review of regularizations. In other words, it is left for the reader to try other canonical transformations within the general scheme presented here.

\section{A General Form of the Hamilton-Jacobi Equation}

We only deal with Hamiltonians of the type
\begin{equation}\label{hamiltonian}
\mathcal{K}(x_0,x,X_0,X;\mu)\equiv\left(X_0+\mathcal{H}\right)\chi,
\end{equation}
where $x=(x_1,\ldots,x_n)$ are coordinates and $X=(X_1,\ldots,X_n)$ conjugate momenta; $x_0$ is the independent variable and $X_0$ its conjugate momenta in the extended phase space formulation, in which we restrict to the manifold $\mathcal{K}=0$; $\mu$ is a vector of parameters, and the Hamiltonian $\mathcal{H}$ as well as the ``regularizing factor'' $\chi$ may depend on all or some of the parameters defining $\mu$:
\begin{equation}
\chi=\chi(x_0,x,X_0,X;\mu),\qquad\mathcal{H}=\mathcal{H}(x_0,x,-,X;\mu).
\end{equation}
A dash in the place of a variable is used to remark that the corresponding variable is not present.
\par

Hamilton equations are
\begin{equation}
\begin{array}{lll}
\displaystyle\frac{{\rm d}x_0}{{\rm d}\tau}=\frac{\partial\mathcal{K}}{\partial{X}_0}, &\quad
\displaystyle\frac{{\rm d}X_0}{{\rm d}\tau}=-\frac{\partial\mathcal{K}}{\partial{x}_0},\\[2ex]
\displaystyle\frac{{\rm d}x_i}{{\rm d}\tau}=\frac{\partial\mathcal{K}}{\partial{X}_i}, &\quad
\displaystyle\frac{{\rm d}X_i}{{\rm d}\tau}=-\frac{\partial\mathcal{K}}{\partial{x}_i}, &\quad  (i=1,\dots,n)
\end{array}
\end{equation}
 where $\tau$ is the evolution parameter of the flow. Note that the first of the previous equations reads
\begin{equation}
\displaystyle\frac{{\rm d}x_0}{{\rm d}\tau}=\chi
\end{equation}
which tells the function $\chi$ ought to verify that $\chi>0$ in its domain. Moreover, in the case of conservative systems $\mathcal{H}=\mathcal{H}(-,x,-,X)$, $X_0$ is an integral, and the manifold $\mathcal{K}=0$ may be also seen as $X_0=-\mathcal{H}=$ constant.
\par

We are interested in canonical transformations 
\begin{equation}
(x_0,x,X_0,X)\stackrel{\mathcal{T}_\Phi}\longrightarrow (y_0,y,Y_0,Y)
\end{equation}
in the sense of Poincar\'e. More precisely, we look for transformations such that they simplify Hamiltonian systems defined by  functions $\mathcal{H}$ which can be written as
\begin{equation}\label{ham1}
\mathcal{H}=\mathcal{H}_0+\mathcal{H}_1
\end{equation}
where $\mathcal{H}_0=\mathcal{H}_0(-,x,-,X)$ defines an integrable system, and $\mathcal{H}_1=\mathcal{H}_1(x_0,x,X)$ is a perturbation. Specifically, we focus on canonical transformations such that the new Hamiltonian $K=K_0+K_1$ satisfies
\begin{equation}\label{HJP}
K_0= \left(X_0+\mathcal{H}_0\right)\chi=\Phi(-,-,Y_0,Y),
\end{equation}
{\sl i.e.} the full reduction of the unperturbed part is carried out. This 
Eq.~(\ref{HJP}) is what we refers as the variant of Poincar\'e to the H-J equation; the classical  case chooses 
$\Phi(-,-,Y_0,Y)\equiv 0$.\par The transformations are defined by 
\begin{equation}\label{ecuatransf}
X_i=\frac{\partial{\cal W}}{\partial x_i}, \qquad y_i=\frac{\partial{\cal W}}{\partial Y_i},\qquad{i}=0,\dots,n
\end{equation}
derived from a generator $\mathcal{W}=\mathcal{W}(x_0,x,Y_0,Y,\mu)$  that is a complete solution of the generalized H-J equation
\begin{equation}\label{parcial}
\left[\frac{\partial\mathcal{W}}{\partial{x}_0}+ \mathcal{H}_0\!\left(x,\frac{\partial\mathcal{W}}{\partial{x}}\right)\right]\,\chi\!\left(x,x_0,\frac{\partial\mathcal{W}}{\partial{x_0}},\frac{\partial\mathcal{W}}{\partial{x}}\right)
=\Phi(Y_0,Y).
\end{equation}\par
Thus, the Hamiltonian $\mathcal{K}$ in the new variables will take the form
\begin{equation}
\mathcal{K}=\mathcal{K}_0+\mathcal{K}_1= \Phi \, + \, \mathcal{H}_1\,\chi
\end{equation}
where $\mathcal{H}_1$ and $\chi$ are expressed in the new variables. Note that  in what follows we take $\mathcal{H}_1=0$. \par
In this paper we limit to  generators of the form
\begin{equation}\label{ecuacion1}
\mathcal{W}=\sum_{0\,\leq\, i\,<\, n-1}\,x_i\,Y_i+\mathcal{R}(x_n,Y_0,Y),
\end{equation}
and regularizing function $\chi=\chi(x_n,X_0,X_1,\dots,X_{n-1})$.
Hence,  from Eq. (\ref{parcial}) we may write
\begin{equation}\label{parcial2}
\mathcal{H}_0\!\left(x,Y_1,\dots,Y_{n-1},\frac{\partial\mathcal{R}}{\partial{x}_n}\right)=
\frac{\Phi(Y_0,Y_1,\dots,Y_{n})}{\chi(x_n,Y_0,Y_1,\dots,Y_{n-1})}-Y_0.
\end{equation}
\par
Depending on the form of $\mathcal{H}_0$, Eq. (\ref{parcial2}) may be solved for $\partial\mathcal{R}/\partial{x}_n$ and, therefore, $\mathcal{R}$ is computed from a quadrature, which solution will depend on the choices made for $\Phi$ and $\chi$. Other possibilities are under investigation \cite{FerrerLara09}. Note that, in fact,  there is no reason why we should impose on $\mathcal{H}_0$ to be cyclic in $x_0$. What we have presented above, properly adapted, remains valid if we lift that constraint. This is referred in the literature as nonautonomuos systems; the driven oscillator, the relativistic particle, etc are just  simple examples within that category (for more recent systems of interest see Struckmeier \cite{Struckmeier}). In the families we will study below the possible presence of $x_0$ occurs in the perturbing part.\par

Apart from the general case of Eq. (\ref{parcial2}), transformations non-based on the homogeneous formalism ($\chi\equiv1$, $\Phi=Y_0+\Psi$), adopt the simpler formulation
\begin{equation}\label{parcial3}
\mathcal{H}_0\!\left(x_n,Y_1,\dots,Y_{n-1},\frac{\partial\mathcal{R}}{\partial{x}_n}\right)=\Psi(Y_1,\dots,Y_{n}).
\end{equation}

In the two types of Hamiltonian systems we study below, instead of $(x_0,x,X_0,X)$ we will start each Section considering briefly the two symplectic charts on which our study relies. In doing so we will use the standard notations, although  there are more than one: Andoyer variables \cite{AndoyerBook} for rotational motion and Whittaker (polar-nodal) variables \cite{Whittaker} for orbital dynamics. 

With respect to the new variables $(y_0,y,Y_0,Y)$, we will use $(d,\gamma,v,u,D,\Gamma,\Upsilon,U)$ for the rotational families and $(f,g,h,u,F,G,H,U)$ for the orbital ones. For the different sets within each family, we do not find necessary to distinguish among them, like using subindices, etc. Each one distinguishes itself by the expression taken by  $\mathcal{K}_0$. 
\section{Rigid body transformations}

Within the frame set up by Eqs (\ref{HJP}) and (\ref{ecuatransf}) we will consider first canonical transformations derived from free rigid body Hamiltonian function. As it is well known, meanwhile Euler of variables continues to be the ones used  in classical and recent books (see \cite{Goldstein2003}, \cite{MarsdenRatiu}) at the same time in research papers is already customary to consider those systems formulated in Andoyer variables. More precisely, the Hamiltonian of the free rigid body  takes the form
\begin{equation}\label{Andoyer0}
\mathcal{H}= \mathcal{H}(-,-,\nu,-,M,N;a_i)=\hbox{$\frac{1}{2}$}\,(a_1\sin^2\nu+a_2\cos^2\nu)\,(M^2-N^2)+\hbox{$\frac{1}{2}$}\,a_3\,N^2.
\end{equation}
where $0<a_3\leq a_2\leq a_1$ are parameters. Then, according to our notation Eq. (\ref{hamiltonian}),
we deal with the Hamiltonian 
\begin{equation}\label{Andoyer}
\mathcal{K}=\left[T\, + \,\hbox{$\frac{1}{2}$}\,(a_1\sin^2\nu+a_2\cos^2\nu)\,(M^2-N^2)+\hbox{$\frac{1}{2}$}\,a_3\,N^2\right]{\chi}(\nu,\Lambda,M,T),
\end{equation}
where $(\lambda,\mu,\nu,\Lambda,M,N)$ are Andoyer variables \cite{AndoyerBook},  and $M^2\ge{N}^2$ making the Hamiltonian, (\ref{Andoyer0}), strictly positive and $\frac{1}{2} \chi \,a_3\,M^2\le\mathcal{H}\le\frac{1}{2}\chi\,a_1\,M^2$. As we said in what follows we consider only the generic case $a_3< a_2< a_1$. We do not need to give more details on those variables here. The interested reader may consult the classical note of Deprit \cite{Deprit1967} or in more geometric terms Fass\`o \cite{Fasso96}.
\par
\subsection{The general structure of the transformation}
We look for canonical transformations
\begin{equation}
(\lambda,\mu,\nu,t,\Lambda,M,N,T)\stackrel{\mathcal{T}_{\Phi}}\longrightarrow(d,\gamma,\upsilon,u,D,\Gamma,\Upsilon,U)
\end{equation}
that convert (\ref{Andoyer}) in a certain function $\Phi(D,\Gamma,\Upsilon,U)$ depending only on momenta.
The transformation will be defined through a function $\mathcal{S}=\mathcal{S}(\lambda,\mu,\nu,t,D,\Gamma,\Upsilon,U)$ in mixed variables such that
\begin{equation}\label{transformation0}
d=\mathcal{S}_D,\quad
\gamma=\mathcal{S}_\Gamma,\quad
\upsilon=\mathcal{S}_\Upsilon,\quad
u=\mathcal{S}_U,\quad
\Lambda=\mathcal{S}_\lambda,\quad
M=\mathcal{S}_\mu,\quad
N=\mathcal{S}_\nu,\quad
T=\mathcal{S}_{t}.
\end{equation}
\noindent
where we use the notation $\mathcal{S}_x=\partial\mathcal{S}/\partial{x}$. Thus, from (\ref{Andoyer}) we set the H-J equation
\begin{equation}\label{HJsolido}
\left[
\hbox{$\frac{1}{2}$}\,(a_1\sin^2\nu+a_2\cos^2\nu)\left(\mathcal{S}_\mu^2-\mathcal{S}_\nu^2\right)+\hbox{$\frac{1}{2}$}\,a_3\,\mathcal{S}_\nu^2+\mathcal{S}_t\right]{\chi}(\nu,\mathcal{S}_\lambda,\mathcal{S}_\mu,\mathcal{S}_{t}),
=\Phi({D},\Gamma,\Upsilon,U)
\end{equation}
and $\mathcal{S}$ is chosen in separate variables as
\begin{equation}
\mathcal{S}= U\,t+\Upsilon\,\lambda+\Gamma\,\mu+\mathcal{W}(\nu,{D},\Gamma,\Upsilon,U)
\end{equation}
because $t$, $\lambda$, and $\nu$ are cyclic in (\ref{Andoyer}). Then,
\begin{equation}
\mathcal{W}=\Gamma\,\int_{\nu_0}^\nu\sqrt{Q(\nu,{D},\Gamma,\Upsilon,U)}\,{\rm d}\nu,
\end{equation}
where
\begin{equation}
Q=\frac{a_1\sin^2\nu+a_2\cos^2\nu-\mathcal{A}}{a_1\sin^2\nu+a_2\cos^2\nu-a_3},
\qquad
\mathcal{A}=\frac{2}{\Gamma^2}\left(\frac{\Phi}{\chi}-U\right),
\end{equation}
and $\sqrt{Q}$ must be real for all $\nu$; therefore, $\Phi<{\chi}\,(\frac{1}{2}a_2\,\Gamma^2+U)$.
\par

If we assume that the regularizing factor depends only on the coordinate $\nu$, the transformation (\ref{transformation0}) are
\begin{eqnarray} \label{delta}
d &=& \frac{1}{\Gamma}\,\Phi_D\,\mathcal{I}_3,\\[0.5ex] \label{gamma}
\gamma &=& \mu+\mathcal{I}_1+\frac{2U}{\Gamma^2}\,\mathcal{I}_2-\left(\frac{2\Phi}{\Gamma^2}-\frac{1}{\Gamma}\,\Phi_\Gamma\right)\mathcal{I}_3,\\[0.5ex] \label{upsilon}
\upsilon &=& \lambda+\frac{1}{\Gamma}\,\Phi_\Upsilon\,\mathcal{I}_3, \\[0.5ex] \label{u}
u &=& t-\frac{1}{\Gamma}\,\mathcal{I}_2+\frac{1}{\Gamma}\,\Phi_U\,\mathcal{I}_3 \\[0.5ex] \label{N}
N &=& \Gamma\,\sqrt{Q}, \\[1.5ex] \label{U}
T &=& U,\qquad \Lambda\,=\,\Upsilon,\qquad M \,=\, \Gamma,
\end{eqnarray}
where
\begin{equation}\label{III123}
\mathcal{I}_1=\int_{\nu_0}^\nu\sqrt{Q}\,{\rm d}\nu, \qquad
\mathcal{I}_2=\int_{\nu_0}^\nu\frac{Q_\mathcal{A}}{\sqrt{Q}}\,{\rm d}\nu, \qquad
\mathcal{I}_3=\int_{\nu_0}^\nu\frac{Q_\mathcal{A}}{{\chi}\,\sqrt{Q}}\,{\rm d}\nu.
\end{equation}

\subsection{The case $\chi=1$}
We only deal with transformations with $\chi=1$. Therefore, $\mathcal{I}_2\equiv\mathcal{I}_3$.
\par

Note that a restriction of Andoyer variables is $|N|\le{M}$, which, because of (\ref{N}), further constrains the values of the new Hamiltonian to be into the positive interval $\Phi\in(\frac{1}{2}a_3\,\Gamma^2,\frac{1}{2}a_2\,\Gamma^2)$. In addition, one should be aware that the sign of $\sqrt{Q}$ must be taken in accordance with the sign of $N$.
\par

Again, the transformation depends on the integration of two quadratures, which closed form solution requires well known changes of variables. Thus, introducing the parameters\footnote{
For the sake of linking with recent literature we used Fukushima's notation \cite{Fukushima2008b}. Deprit \cite{DepritElipe1993} calls $\delta\equiv\alpha_3$, $\epsilon\equiv\alpha_3^0$, and $m\equiv{k}_3^2$, and Sadov \cite{Sadov} names $m\equiv\lambda$ and $f=\kappa^2$.}
\begin{equation}\label{a30}
\epsilon=\frac{a_1-a_2}{a_1-a_3},\qquad f=\frac{a_1-a_2}{a_2-a_3},
\end{equation}
and the functions $\delta=\delta(D,\Gamma,\Upsilon,U;a_1,a_2,a_3)$, $m=m(\delta)$, defined by
\begin{eqnarray}
\delta &=& \frac{a_1-a_2}{\Gamma^2\,a_1-2(\Phi-U)}\,\Gamma^2,\\
m &=& \frac{\delta-\epsilon}{1-\epsilon}=\frac{a_1-a_2}{\Gamma^2\,a_1-2(\Phi-U)}\,\frac{2(\Phi-U)-\Gamma^2\,a_3}{a_2-a_3},
\end{eqnarray}
the quadratures in (\ref{III123}) are solved to give
\begin{equation}
\mathcal{I}_1=\frac{m\,F(\psi|m)-(m+f)\,\Pi(\psi,-f|m)}{\sqrt{\delta\,f}},
\end{equation}
\begin{equation}
\mathcal{I}_2=\frac{\Gamma\,F(\psi|m)}{\sqrt{a_2-a_3}\,\sqrt{\Gamma^2\,a_1-2(\Phi-U)}}.
\end{equation}
where $F(\psi|m)$ is the elliptic integral of the first kind of modulus $m$ which amplitude $\psi$ is unambiguously defined through
\begin{equation}\label{Deprit2Serretnu}
\cos\nu=\frac{\sin\psi}{\sqrt{1-\epsilon\cos^2\psi}},\quad\sin\nu=\frac{\sqrt{1-\epsilon}\,\cos\psi}{\sqrt{1-\epsilon\cos^2\psi}},\qquad
{\rm d}\nu=-\frac{\sqrt{1-\epsilon}}{1-\epsilon\cos^2\psi}\,{\rm d}\psi.
\end{equation}
On the other hand, $\Pi(\psi,-f;m)$ is the elliptic integral of the third kind of modulus $m$, characteristic $-f$, and amplitude $\psi$. Alternatively, one can write
\begin{equation}\label{I1}
\mathcal{I}_1=\sqrt{\frac{f}{\delta}}\,(1-\delta)\,\Pi(\phi,\delta|m),
\end{equation}
where the new amplitude $\phi$ is unambiguously defined through
\begin{equation}\label{Deprit2Serretnufi}
\cos\nu=\frac{\cos\phi}{\sqrt{1-\delta\sin^2\phi}},\qquad
\sin\nu=\frac{\sqrt{1-\delta}\,\sin\phi}{\sqrt{1-\delta\sin^2\phi}},\qquad
{\rm d}\nu=\frac{\sqrt{1-\delta}}{1-\delta\sin^2\phi}\,{\rm d}\phi.
\end{equation}
\par

If we further introduce 
\begin{equation}\label{sigma}
\sigma=\frac{\Gamma}{\sqrt{\Gamma^2a_1-2(\Phi-U)}\,\sqrt{a_2-a_3}}=\frac{\sqrt{\delta\,f}}{a_1-a_2}=\frac{1}{a_2-a_3}\,\sqrt{\frac{\delta}{f}},
\end{equation}
then, the transformation equations above, (\ref{delta})--(\ref{N}), adopt the compact form
\begin{eqnarray} \label{delta0}
d &=& \hspace{0.7cm}\frac{\sigma}{\Gamma}\,\Phi_D\,F(\psi,m)  \\[0.5ex]
\gamma &=&\mu-\frac{\sigma}{\Gamma}\,\Big[\left(\Gamma\,a_3-\Phi_\Gamma\right)F(\psi,m)+\Gamma\,(a_1-a_3)\,\Pi(\psi,-f|m)\Big] \\
&=&\mu-\frac{\sigma}{\Gamma}\left[\!
\left(\!2\frac{\Phi-U}{\Gamma}-\Phi_\Gamma\right)F(\psi,m)-\left(2\frac{\Phi-U}{\Gamma}-\Gamma\,a_2\right)\Pi(\phi,\delta|m)\right]
\\[0.5ex]
\upsilon &=& \lambda+\frac{\sigma}{\Gamma}\,\Phi_\Upsilon\,F(\psi,m)= \lambda+\frac{\Phi_\Upsilon}{\Phi_D}\,d
\\[0.5ex] \label{urigid}
u &=& t+\frac{\sigma}{\Gamma}\left(\Phi_U-1\right)F(\psi|m)
=t+\frac{\Phi_U-1}{\Phi_D}\,d, \\[0.5ex]
\label{N0}
N &=& \Gamma\,\sqrt{\frac{\epsilon\,(1-\delta\cos^2\nu)}{\delta\,(1-\epsilon\cos^2\nu)}}
=\Gamma\,\sqrt{\frac{\epsilon}{\delta}\,(1-m\sin^2\psi)}, 
\end{eqnarray}
where we maintain two expressions for $\gamma$ to ease comparisons with transformations in the literature. Remark that $\Phi$ remains arbitrary. Therefore, (\ref{sigma})--(\ref{N0}) define a family of canonical transformations that provide complete reduction of the Euler-Poinsot problem. Note, besides, from (\ref{urigid}) that reduced Hamiltonians of the form $\Phi=U+\Psi(D,\Gamma,\Upsilon)$ are required for transformations preserving the time scale.
\par

Despite the family of transformations above has been obtained from a specific dynamical system, the rigid-body problem, once it has been derived it may be applied to any problem we wish. However, successful transformations for different problems are closely dependent on the selection of $\Phi$. Thus, for instance, if we are to apply one of the transformations of the family (\ref{delta0})--(\ref{N0}) above to reduce the rigid body dynamics or to study a perturbed rigid body, we should consider specific facts. Because neither $\Lambda$ nor $T$ are altered by the transformation it seems natural to choose $\Phi=U+\Psi(D,\Gamma)$. Then, $u$ is the time, $\upsilon$, $\Upsilon$, $U$, $\Gamma$ and $D$, remain constant, and
\begin{equation}
\gamma=\gamma_0+\Phi_\Gamma\,u,\qquad
d=d_0+\Phi_D\,u.
\end{equation}
\par

Because the super-integrability of the Euler-Poinsot problem the closure of a generic trajectory is a two-torus \cite{Benettin}, Therefore, $\Psi$ should depend both on $D$ and $\Gamma$ if we want to retain the topology of the problem. Specific selections of $\Psi$ depending only on $D$ ---whose appearance is mandatory for equation (\ref{delta0}) to be defined--- will produce just periodic solutions. Even particular selections $\Psi=\Psi(D,\Gamma)$ may reduce the solutions to just periodic orbits. For instance, if we impose 
\begin{equation}
i\,\Psi_\Gamma=j\,\Psi_D,
\end{equation}
$i$, $j$, integer, any computed $\Phi=U+\Psi(i\,D+j\,\Gamma)$ with arbitrary $\Psi$ will provide only periodic solutions.
\par

We discuss here some other possible choices of $\Phi$. Without any application in mind, one criteria might be `to simplify' the coefficients of the variables in the transformation.
\par

Thus, for instance, $\sigma=D$ in (\ref{sigma}) implies
\begin{equation}
\Phi=U+\frac{\Gamma^2}{2}\left(a_1-\frac{D^2}{a_2-a_3}\right)
\end{equation}
where $\Gamma$ has dimensions of momentum and $D$ of inverse of momentum of inertia. This Hamiltonian simplifies the transformation equations, (\ref{delta0})--(\ref{N0}), to
\begin{eqnarray}
d &=& \hspace{0.35cm}-\frac{\Gamma\,D^2}{a_2-a_3}\,F(\psi,m)  \\
\gamma &=& \mu+D\left(a_1-a_2-\frac{D^2}{a_2-a_3}\right)\Pi(\phi,\delta|m) \\
\upsilon &=& \lambda \\
u &=& t, \\
N &=& \Gamma\,D\,\sqrt{\frac{1-m\sin^2\psi}{(a_1-a_3)\,(a_2-a_3)}},
\end{eqnarray}
\par

To discuss other simplifications, we find it convenient to introduce
\begin{equation}\label{n1n2n3}
s_1=\frac{\Gamma}{\sigma\,\Phi_D},\qquad
s_2=\left(2\frac{\Phi-U}{\Gamma}-\Phi_\Gamma\right)\frac{1}{\Phi_D},\qquad
s_3=\frac{\Phi_\Upsilon}{\Phi_D},
\end{equation}
and consider the possible choices of $\Phi$ that make $s_i$ ($i=1,2,3$) independent of the momenta.

\subsubsection{Case $s_1=s_{1,0}$ and $s_3=s_{3,0}$ constant.}\label{n1cte}

Then, from the first and third of (\ref{n1n2n3}) and from (\ref{sigma}), we get
\begin{equation}
\Phi_\Upsilon=s_{3,0}\,\Phi_D=\frac{s_{3,0}}{s_{1,0}}\,\sqrt{a_2-a_3}\,\sqrt{a_1\,\Gamma^2-2(\Phi-U)},
\end{equation}
and, therefore
\begin{equation} 
\Phi-U=\frac{a_1}{2}\,\Gamma^2-\frac{a_2-a_3}{2\,s_{1,0}^2}\left[s_{3,0}\,\Upsilon+\Psi(\Gamma,D,U)\right]^2.
\end{equation}
with $\Psi$ arbitrary.
\par

Because the topology of the problem, we can add the condition $\Phi_\Upsilon=0$, and therefore $s_{3,0}\equiv0$. In addition, if we want to preserve the time scale (\ref{urigid}), we restrict to the case
\begin{equation} 
\Phi-U=\frac{a_1}{2}\,\Gamma^2-\frac{a_2-a_3}{2\,s_{1,0}^2}\,\Psi(\Gamma,D)^2.
\end{equation}
\par

\paragraph{Sadov's transformation.}
The choice
\[
\Psi\equiv s_{1,0}\,\sqrt{\frac{a_1-a_3}{a_2-a_3}}\,D
\]
gives
\begin{equation}\label{SadovHam}
\Phi-U=\frac{a_1\,\Gamma^2}{2}\left[1-\frac{a_1-a_3}{a_1}\left(\frac{{D}}{\Gamma}\right)^2\right]
=\frac{a_1\,\Gamma^2}{2}\left[1-\frac{a_1-a_3}{a_1}\frac{\kappa^2}{\kappa^2+\lambda^2}\right],
\end{equation}
that is Sadov's \cite{Sadov} reduced Hamiltonian in which $\kappa^2\equiv{f}$ and
\begin{equation}
\lambda^2=\kappa^2\left(\frac{\Gamma^2}{{D}^2}-1\right).
\end{equation}
\par

\paragraph{Kinoshita's case.}
If we instead choose
\begin{equation}\label{s10Kinoshita}
\Psi\equiv s_{1,0}\,\sqrt{\frac{a_1\,(a_1+a_2-2a_3)}{(a_2-a_3)\,(a_1+a_2)}}\,D,
\end{equation}
we find 
\begin{equation}\label{KinoshitaHam}
\Phi-U=\frac{2a_1}{a_1+a_2}\,\tilde{\mathcal{H}},\qquad
\tilde{\mathcal{H}}=\frac{1}{2}\left(\frac{a_1+a_2}{2}\,\Gamma^2+\frac{1}{b}\,{D}^2\right),\quad
\frac{1}{b}=a_3-\frac{1}{2}\,(a_1+a_2)<0,
\end{equation}
where $\tilde{\mathcal{H}}$ is Kinoshita's reduced Hamiltonian \cite[see Eq. (10') on p. 433]{Kinoshita72}.
Alternatively, the scaling may be avoided in our general formulation by choosing a regularizing factor $\chi=2a_1/(a_1+a_2)$ independent of the variables.

Thus, the major objection to Sadov's and Kinoshita's elements of being related to Andoyer canonical variables through implicit transformations is easily circumvented using the general transformation (\ref{sigma})--(\ref{N0}) above particularized for either Sadov's or Kinoshita's Hamiltonians, Eq. (\ref{SadovHam}) and (\ref{KinoshitaHam}) respectively. 
\par

\paragraph{Fukushima proposal.}
If we choose now
\begin{equation}\label{s10Fukushima}
\Psi\equiv s_{1,0}\,\sqrt{\frac{a_1}{a_2-a_3}}\,\sqrt{\Gamma^2-D^2},
\end{equation}
we get Fukushima's proposal \cite{Fukushima2008b}
\begin{equation}
\Phi=U+\frac{1}{2}\,a_1\,{D}^2,
\end{equation}
that, depending on less than the required momenta, constrain the topology of the transformed Hamiltonian to periodic solutions only. Nevertheless, this way of proceeding may be perfectly adequate for a perturbation theory, as shown in \cite{Yanguas}.

\paragraph{TR-type mapping.} In a similar way to Fukushima, we propose to choose
\begin{equation}\label{s10Fukushima}
\Psi\equiv s_{1,0}\,\sqrt{\frac{a_1}{a_2-a_3}}\,(\Gamma-D),
\end{equation}
leading to
\begin{equation}
\Phi-U=a_1\,D\left(\Gamma-\mbox{$\frac{1}{2}$}D\right)
\end{equation}
that is formally equal to the reduced Hamiltonian of the TR-mapping \cite{Scheifele,Deprit1981a} given in Eq. (\ref{HillTR}) below. It is quadratic, depends on two momenta ---as it should be for the more general case of a super-integrable three degrees of freedom problem with four independent integrals \cite{Benettin}--- and the momenta keep the dimensions of the original problem because the Hamiltonian is multiplied by the inverse of the momentum of inertia $a_1$.
\par

Then, the reduced system has four elements $(\upsilon,D,\Gamma,\Upsilon)$ and two variables that evolve linearly with time
\[
\gamma=\gamma_0+a_1\,D\,t,\qquad d=d_0+a_1\,(\Gamma-D)\,t,
\]
the condition $\Gamma/D$ rational providing the subset of periodic solutions.

\subsubsection{Case $s_2=s_{2,0}$ and $s_3=s_{3,0}$ constant}\label{n2cte}
From the second and third of (\ref{n1n2n3}) we get
\begin{equation}\label{s20}
s_{2,0}\,\Phi_D=\frac{s_{2,0}}{s_{3,0}}\,\Phi_\Upsilon=2\,\frac{\Phi-U}{\Gamma}-\Phi_\Gamma
\quad\Rightarrow\quad
\Phi=U+\Gamma^2\,\Psi({D}-s_{2,0}\,\Gamma+s_{3,0}\,\Upsilon,U)
\end{equation}
where $\Psi$ is arbitrary.
\par

A simple choice satisfying the condition in (\ref{s20}) could be $\Phi=U+\Gamma^2\,({D}-s_{2,0}\,\Gamma)$, for instance. Particular cases in the literature make $s_{2,0}=0$. Thus, Deprit and Elipe \cite{DepritElipe1993} choose
\begin{equation}\label{DepritHam}
\Phi-U=\frac{1}{2}\,\frac{\Gamma^2}{D},
\end{equation}
Other solutions could be, for instance,
\begin{equation}
\Phi-U=\Gamma^2\,{D}^2,\qquad
\Phi-U=\Gamma^2\,{D}.
\end{equation}

\section{Orbital Canonical transformations}

What we have said at the beginning of the previous Section for canonical variables in rotational dynamics, it appears again in orbital variables.  Thus, meanwhile Hamiltonian particle dynamics still happens to be introduced in  spherical canonical variables (see for instance \cite{BoccalettiPucacco}, \cite{Goldstein2003}, \cite{JoseSaletan},  \cite{SussmanWisdom}), at the same time in space research  another canonical set of variables is widely in use; we refer to nodal-polar variables,  already considered for planetary theories almost a century ago \cite{Hill}. The main difference of both set of variables is that nodal-polar variables carries out a double reduction (axial and central symmetries) of Kepler type systems, meanwhile spherical variables only deals with the axial symmetry. \par

As in the search of orbital canonical transformations we build up our H-J equation based on nodal-polar variables, we briefly give a description about them. Taking a reference frame $(O,{\bf e}_1,{\bf e}_2,{\bf e}_3)$, they are introduced in a natural way associated with the ``instantaneous plane of motion'' whose characteristic vector is ${\bf x}\times {\bf X}$. The nodal-polar variables $(r,\nu,\theta)$ are defined as follows: the variable $r$ is $\|{\bf x}\|$; the angle $\nu$ called the {\sl ascending node}, is defined by $\cos \nu = {\bf e}_1\cdot \ell$ and $\sin \nu = {\bf e}_2\cdot \ell$, where $\ell$ is the unit vector defined by ${\bf e}_3\times ({\bf x}\times {\bf X})$; the variable $\theta$ is a polar angle in the plane of motion giving the position of ${\bf x}$ reckoned from the vector $\ell$. The variables $(R,N,\Theta)$ are the corresponding conjugate momenta. For more details and explicit expressions of the transformation $(x,y,z,X,Y,Z) 
 \rightarrow (r,\nu,  \theta, R, N,\Theta)$, called by some authors Whittaker transformation, see for instance \cite{Deprit1970}.

\subsection{The general structure of the transformations}
Given the function
\begin{equation}\label{KeplerHill}
\mathcal{H}=\left[\frac{1}{2}\left(R^2+\frac{\Theta^2}{r^2}\right)+\mathcal{V}(r)+T\right]\chi(r,\Theta,N,T)
\end{equation}
where $(r,\theta,\nu,t,R,\Theta,N,T)$ are Hill or polar-nodal variables in the extended phase space and $\mathcal{V}$ only depends on distance, we look for a canonical transformation
\[
(r,\theta,\nu,t,R,\Theta,N,T)\stackrel{\mathcal{T}_{\Phi}}\longrightarrow (f,g,h,u,F,G,H,U)
\]
that converts (\ref{KeplerHill}) in a certain function $\Phi(F,G,H,U)$ depending only on the momenta.
\par

The transformation will be defined through a generating function in mixed variables $\mathcal{S}=\mathcal{S}(r,\theta,\nu,t,F,G,H,U)$ such that
\begin{equation}\label{transformationHill}
f=\mathcal{S}_F,\quad
g=\mathcal{S}_G,\quad
h=\mathcal{S}_H,\quad
u=\mathcal{S}_U,\quad
R=\mathcal{S}_r,\quad
\Theta=\mathcal{S}_\theta,\quad
N=\mathcal{S}_\nu,\quad
T=\mathcal{S}_t.
\end{equation}
Then, from (\ref{KeplerHill}) we set the H-J equation
\begin{equation}\label{HJkepler}
\left[\frac{1}{2}\left(\mathcal{S}_r^2+\frac{1}{r^2}\,\mathcal{S}_\theta^2\right)+\mathcal{V}(r)+\mathcal{S}_t\right]
\chi(r,\mathcal{S}_\theta,\mathcal{S}_\nu,\mathcal{S}_t)
=\Phi(F,G,H,U)
\end{equation}

Because $t$, $\theta$, and $\nu$ are not present in (\ref{KeplerHill}), the generating function may be chosen in separate variables
\begin{equation}
\mathcal{S}=U\,t+H\,\nu+G\,\theta+\mathcal{W}(r,F,G,H,U)
\end{equation}
Then,
\begin{equation}\label{HJkepler000}
\frac{1}{2}\left(\mathcal{W}_r^2+\frac{G^2}{r^2}\right)+\mathcal{V}(r)+U=
\frac{\Phi(F,G,H,U)}{\chi(r,G,H,U)}
\end{equation}
that can be solved for $\mathcal{W}$ by a simple quadrature: 
\begin{equation}
\mathcal{W}=\int_{r_0}^r\sqrt{Q(r,F,G,H,U)}\,{\rm d}r,
\end{equation}
where
\begin{equation}\label{qu}
Q=2\frac{\Phi(F,G,H,U)}{\chi\left(r,G,H,U\right)}-2\mathcal{V}(r)-\frac{G^2}{r^2}-2U\ge0.
\end{equation}
Therefore, the transformation is
\begin{eqnarray} \label{efemu0}
f &=& \Phi_F\,\mathcal{I}_3, \\[0.5ex] \label{gemu0}
g &=& \theta+G\,\mathcal{I}_1+\Phi_G\,\mathcal{I}_3-\Phi\,\int_{r_0}^r\,\frac{\chi_G}{\chi^2\,\sqrt{Q}}\,{\rm d}r, \\[0.5ex] \label{hachemu0}
h &=& \nu+\Phi_H\,\mathcal{I}_3-\Phi\,\int_{r_0}^r\,\frac{\chi_H}{\chi^2\,\sqrt{Q}}\,{\rm d}r, \\[0.5ex] \label{umu0}
u &=& t-\mathcal{I}_2+\Phi_U\,\mathcal{I}_3-\Phi\,\int_{r_0}^r\,\frac{\chi_U}{\chi^2\,\sqrt{Q}}\,{\rm d}r, \\[0.6ex] \label{Erremu}
R &=& \mathcal{W}_r =\sqrt{Q}, \\[1.5ex] \label{GET}
\Theta &=& G,\qquad N\,=\, H,\qquad T\,=\,U.
\end{eqnarray}
where
\begin{equation}\label{Hillquadratures}
\mathcal{I}_1=\int_{r_0}^r\frac{1}{\sqrt{Q}}\,{\rm d}\!\left(\frac{1}{r}\right),\qquad
\mathcal{I}_2=\int_{r_0}^r\frac{{\rm d}r}{\sqrt{Q}},\qquad
\mathcal{I}_3=\int_{r_0}^r\frac{{\rm d}r}{\chi\,\sqrt{Q}}.
\end{equation}

\paragraph{The case $\chi=\chi(r)$. }
Transformations are clearly simplified when we choose the regularizing factor $\chi$ independent of the momenta $G$, $H$, and $U$. Thus, if $\chi=\chi(r)$, we have
\begin{equation}
\chi_G=\chi_H=\chi_U=0,
\end{equation}
and the rightmost terms on the right side of (\ref{efemu0})--(\ref{umu0}) vanish, which turn these equations into
\begin{eqnarray} \label{efemu}
f &=& \Phi_F\,\mathcal{I}_3, \\ \label{gemu}
g &=& \theta+G\,\mathcal{I}_1+\Phi_G\,\mathcal{I}_3, \\ \label{hachemu}
h &=& \nu+\Phi_H\,\mathcal{I}_3, \\ \label{umu}
u &=& t-\mathcal{I}_2+\Phi_U\,\mathcal{I}_3.
\end{eqnarray}

Note that $\chi=1$ and $\chi=r^2$ make $\mathcal{I}_2=\mathcal{I}_3$ and $\mathcal{I}_3=-\mathcal{I}_1$, respectively.  We plan to discuss these choices in detail below, focusing on a two-parametric family $\mathcal{V}=\mathcal{V}(r;\mu,b)$ of radial potentials.
\par
\subsection{Families of the isochronal potential}
In what follows we only deal with the transformation (\ref{efemu})--(\ref{umu}) and, in addition, limit our study to the specific case of H\'enon's isochronal potential \cite{Henon,BoccalettiPucacco}
\begin{equation}\label{isochronal}
\mathcal{V}=-\frac{\mu}{b+\sqrt{b^2+r^2}},
\end{equation}
where $b$ and $\mu$ are parameters. A relevant, particular case of the isochronal is the Keplerian potential $\mathcal{V}=-\mu/r$ that occurs for $b=0$.
\par

For the isochronal potential (\ref{isochronal}) we find convenient to write (\ref{qu}) as
\begin{equation}\label{ququ}
Q=\frac{r^2+b^2}{r^2}\,\mathcal{Q}\ge0,\qquad\mathcal{Q}=-\alpha\left(\frac{p}{r^2+b^2}-\frac{2}{\sqrt{r^2+b^2}}+\frac{1}{a}\right),
\end{equation}
where $\alpha$, $p$, and $a$, are certain functions of the momenta and parameters, which only will be determined after the regularizing parameter $\chi$ has been chosen. Then,
\begin{equation}
\mathcal{Q}=\alpha\,p\left(\frac{1}{s}-\frac{1}{s_1}\right)\left(\frac{1}{s_2}-\frac{1}{s}\right),
\end{equation}
where $s_1\ge(s=\sqrt{r^2+b^2})\ge{s}_2$ are the two possible roots of the conic $\mathcal{Q}=0$:
\begin{equation}\label{r12}
s_{1,2}=\frac{p}{1\pm\sqrt{1-p/a}}=\frac{a\,(1-e^2)}{1\pm{e}}=a\,(1\pm{e}),\qquad e^2=1-\frac{p}{a}<1.
\end{equation}
\subsubsection{Introducing auxiliary variables}
The roots $s_{1,2}$, the extreme values of $s$, make natural the introduction of the {\sl auxiliary variables} $\psi$ and $\phi$, defined through
\begin{equation}\label{reccentric}
\sqrt{r^2+b^2}=a\,(1-e\cos\psi),\qquad{\rm d}\sqrt{r^2+b^2}=a\,e\sin\psi\,{\rm d}\psi,
\end{equation}
\begin{equation}\label{rtrue}
\sqrt{r^2+b^2}=\frac{p}{1+e\cos\phi},\qquad{\rm d}\!\left(\frac{1}{\sqrt{r^2+b^2}}\right)=-\frac{e}{p}\sin\phi\,{\rm d}\phi,
\end{equation}
in order to ease the solution of quadratures in (\ref{Hillquadratures}), which we find convenient to write
\begin{eqnarray} \label{I1isochronal}
\mathcal{I}_1 &=& \int_{r_0}^r\frac{r^2+b^2}{r^2\,\sqrt{\mathcal{Q}}}\,{\rm d}\left(\frac{1}{\sqrt{r^2+b^2}}\right),\\ \label{I2isochronal}
\mathcal{I}_2 &=& \int_{r_0}^r\frac{1}{\sqrt{\mathcal{Q}}}{\rm d}\left(\sqrt{r^2+b^2}\right),\\ \label{I3isochronal}
\mathcal{I}_3 &=& -\int_{r_0}^r\frac{r^2+b^2}{\chi\,\sqrt{\mathcal{Q}}}\,{\rm d}\left(\frac{1}{\sqrt{r^2+b^2}}\right)=\int_{r_0}^r\frac{1}{\chi\,\sqrt{\mathcal{Q}}}{\rm d}\left(\sqrt{r^2+b^2}\right).
\end{eqnarray}
\par

For given values of the regularizing factor $\chi$, the quadratures above, (\ref{I1isochronal})--(\ref{I3isochronal}), may be integrated in closed form without need of specifying the new Hamiltonian, which remains as a formal function of the momenta $\Phi=\Phi(F,G,H,U)$. 
\par

Details on the families of canonical transformations generated by the cases $\chi=1$, $\chi=b+\sqrt{b^2+r^2}$, and $\chi=r^2$ are given below. Besides, for the Keplerian case $b=0$ we recover Delaunay's \cite{Delaunay}, Levi-Civita's \cite{LeviCivita1,LeviCivita2}, and Hill's \cite{Hill} canonical transformations, as particular examples of our families. An elegant alternative for the derivation of these historic transformations may be found in \cite{Andoyer1913}.

\subsection{Case I: Delaunay's family of transformations $\chi=1$.}

From (\ref{qu}) and (\ref{ququ}) we obtain
\begin{equation}\label{qudelaunay}
\alpha=\mu,\qquad
a=\frac{\mu}{2(U-\Phi)},\qquad
p=\frac{G^2}{\mu}+2b-\frac{b^2}{a}.
\end{equation}
Then, introducing the change of (\ref{reccentric}) in (\ref{I2isochronal}), $\mathcal{I}_2$ is easily solved to give
\begin{equation}\label{I2delaunay}
\mathcal{I}_2=\mathcal{I}_3=\frac{\mu}{\sqrt{8(U-\Phi)^3}}\,(\psi-e\sin\psi).
\end{equation}
Analogously, the change of (\ref{rtrue}) is introduced in (\ref{I1isochronal}) to solve the quadrature $\mathcal{I}_1$. We obtain
\begin{equation}\label{I1delaunay}
\mathcal{I}_1=-\frac{\phi_1}{2G}-\frac{\phi_2}{2\sqrt{G^2+4b\,\mu}},
\end{equation}
where  we have introduced two new {\sl auxiliary variables} $\phi_1$, $\phi_2$, defined by means of the trigonometric relations \cite{Yanguas}
\begin{eqnarray}\label{phi1}
\tan\frac{\phi_1}{2} &=& \sqrt{\frac{1+e-b/a}{1-e-b/a}}\,\sqrt{\frac{1-e}{1+e}}\tan\frac{\phi}{2}
=\sqrt{\frac{1+e-b/a}{1-e-b/a}}\tan\frac{\psi}{2},
\\ \label{phi2}
\tan\frac{\phi_2}{2} &=& \sqrt{\frac{1+e+b/a}{1-e+b/a}}\,\sqrt{\frac{1-e}{1+e}}\tan\frac{\phi}{2}
=\sqrt{\frac{1+e+b/a}{1-e+b/a}}\tan\frac{\psi}{2}.
\end{eqnarray}
\par

Substitution of $\mathcal{I}_1$ and $\mathcal{I}_2\equiv\mathcal{I}_3$, given by (\ref{I1delaunay}) and (\ref{I2delaunay}), respectively, in (\ref{efemu})--(\ref{umu}) provides the family of canonical transformations of the isochronal potential to which we attach the name of Delaunay.
\par

Among the variety of possible choices of the new Hamiltonian $\Phi$, a simplifying option is to take $\Phi=U+\Psi(F,G,H;\mu,b)$, that makes $u$ to remain as the original  time. In addition, the choice $\Psi_H=0$, $\Psi_F=\Psi_G=\mu^{-1}\,(-2\Psi)^{3/2}$, may be solved for $\Psi$, to give
\begin{equation}\label{noYanguas}
\Phi=U-\frac{1}{2}\,\frac{\mu^2}{(F+G)^2}
\end{equation}
that maximally simplifies the transformation equations:
\begin{eqnarray} \label{efemu2}
f &=& \psi-e\sin\psi, \\ \label{gemu2}
g &=& \theta-\frac{\phi_1}{2}-\frac{G}{\sqrt{G^2+4b\,\mu}}\,\frac{\phi_2}{2}+f, \\ \label{hachemu2}
h &=& \nu, \\ \label{umu2}
u &=& t,
\end{eqnarray}
while retaining the topology of the original Hamiltonian. Delaunay's selection
 $\Psi_H=0$, $\Psi_G=0$ leads to
\begin{equation}\label{Delaunay}
\Phi=U-\frac{\mu^2}{2F^2}
\end{equation}
that depends on fewer momenta and, therefore, constrains the topology of the original system to periodic solutions only ---which may be adequate for a perturbation study like \cite{Yanguas}.
\par

In the Keplerian case $b=0$, $\mathcal{V}=-\mu/r$, the selection of the new Hamiltonian $\Phi=\Phi(U,F)$ does not constrain the range of solutions and the specific selection of (\ref{Delaunay}) provides the popular Delaunay transformation that, taking into account that $\phi_1=\phi_2=\phi$, is
\begin{equation}
f=\psi-e\sin\psi,\qquad g=\theta-\phi,\qquad h=\nu,\qquad u=t,
\end{equation}
where the most extended notation writes $L\equiv{F}$, $\ell\equiv{f}$.
\par

\subsection{Case II: Levi-Civita's family of transformations~\hbox{$\chi=b+\sqrt{b^2+r^2}$}}
Now, we write
\begin{equation}
\alpha=\Phi+\mu,\qquad
a=\frac{\Phi+\mu}{2U},\qquad
p=\frac{G^2}{\Phi+\mu}+2b-\frac{b^2}{a}.
\end{equation}

The change (\ref{rtrue}) is used to integrate $\mathcal{I}_1$, (\ref{I1isochronal}), and the change (\ref{reccentric}) to integrate $\mathcal{I}_2$ and $\mathcal{I}_3$, (\ref{I2isochronal}) and (\ref{I3isochronal}). We get
\begin{eqnarray} \label{I1LeviCivita}
\mathcal{I}_1 &=& -\frac{\phi_1}{2G}-\frac{\phi_2}{2\sqrt{G^2+4b\,(\mu+\Phi)}},\\ \label{I2LeviCivita}
\mathcal{I}_2 &=& \frac{\mu+\Phi}{\sqrt{8U^3}}\,(\psi-e\sin\psi),\\ \label{I3LeviCivita}
\mathcal{I}_3 &=& \frac{\psi}{\sqrt{2U}}-\frac{b\,\phi_2}{\sqrt{G^2+4b\,(\mu+\Phi)}}.
\end{eqnarray}
where $\phi_1$, $\phi_2$, are the same auxiliary variables defined in (\ref{phi1}) and (\ref{phi2}) respectively. Then, substitution of $\mathcal{I}_1$, $\mathcal{I}_2$, $\mathcal{I}_3$, (\ref{I1LeviCivita})--(\ref{I3LeviCivita}), in (\ref{efemu})--(\ref{umu}) provides the family of canonical transformations of the isochronal potential to which we tie the name of Levi-Civita.
\par

The, Keplerian case $b=0$, $\phi_1=\phi_2=\phi$, gives
\begin{eqnarray} \label{efe2}
f &=& \Phi_F\,\frac{1}{\sqrt{2U}}\,\psi, \\ \label{ge2}
g &=& \theta-\phi+\Phi_G\,\frac{1}{\sqrt{2U}}\,\psi, \\ \label{hache2}
h &=& \nu+\Phi_H\,\frac{1}{\sqrt{2U}}\,\psi, \\ \label{u2}
u &=& t-\frac{\mu+\Phi}{\sqrt{8U^3}}\,(\psi-e\sin\psi)+\Phi_U\,\frac{1}{\sqrt{2U}}\,\psi.
\end{eqnarray}
The canonical transformation still remains undefined. We note in (\ref{efe2}) that $f=\psi$ when $\Phi_F=\sqrt{2U}$, a partial differential equation that may be solved to give
\begin{equation}\label{PhiF}
\Phi=\sqrt{2U}\,\Big[F+\mathcal{C}_1(G,H,U)\Big],
\end{equation}
where $\mathcal{C}_1$ is an arbitrary function of $G$, $H$, and $U$. The simple choice $\mathcal{C}_1=\mu$ provides the ``first'' Levi-Civita \cite{LeviCivita1} transformation\footnote{Note that Levi-Civita proceeds by  scaling $r$ by $1/\sqrt{2U}$, or choosing $\chi=r/\sqrt{2U}$ in our notation. This is equivalent to scaling our new Hamiltonian by the same factor, and yields to Levi-Civita's original Hamiltonian $\Phi=F$ while the transformation (\ref{LeviCivitaOne}) remain identical}
\begin{equation}\label{LeviCivitaOne}
f=\psi,\qquad
g=\theta-\phi,\qquad
h=\nu,\qquad
u=t-\frac{\mu}{\sqrt{8U^3}}\,(f-e\sin{f}).
\end{equation}
\par

Other possibility is to force $\Phi_U=(\mu+\Phi)/(2U)$ so that only periodic oscillations are introduced in the time scale (\ref{u2}). Then,
\begin{equation}\label{PhiU}
\Phi=-\mu+\sqrt{U}\,\mathcal{C}_2(F,G,H).
\end{equation}
with $\mathcal{C}_2$ arbitrary. The combination of both conditions (\ref{PhiF}) and (\ref{PhiU}) leads to
\begin{equation}
\Phi=\sqrt{2U}\left(F-\frac{\mu}{\sqrt{2U}}+\mathcal{C}(G,H)\right),
\end{equation}
with $\mathcal{C}$ arbitrary. If we further take $\mathcal{C}\equiv0$, we get
\begin{equation}
f=\psi,\qquad
g=\theta-\phi,\qquad
h=\nu,\qquad
u=t+\frac{F}{2U}\,e\,\sin{f}.
\end{equation}
that is the famous ``second'' Levi-Civita \cite{LeviCivita2} transformation.

\subsection{Case III: Hill's family of transformations: $\chi=r^2$}

We get now
\begin{equation}
\alpha=\mu,\qquad
a=\frac{\mu}{2U},\qquad{p}=\frac{G^2-2\Phi}{\mu}+2b-\frac{b^2}{a}.
\end{equation}
The required quadratures are solved with the same changes of variables as before, giving
\begin{eqnarray}
\mathcal{I}_1 &=& -\mathcal{I}_3=-\frac{\phi_1}{2\sqrt{G^2-2\Phi}}-\frac{\phi_2}{2\sqrt{G^2-2\Phi+4b\,\mu}},\\
\mathcal{I}_2 &=& \frac{\mu}{\sqrt{8U^3}}\,(\psi-e\sin\psi),
\end{eqnarray}
which substitution in (\ref{efemu})--(\ref{umu}) provides the family of canonical trans\-for\-ma\-tions of the isochronal potential to which we assign the name of Hill.
\par

Focusing on the Keplerian case $b=0$, $\phi_1=\phi_2=\phi$, a simplification choice that makes $f=\phi$ is to select $\Phi$ in such a way that $\Phi_F\,\mathcal{I}_1=-\phi$, then
\begin{equation}
\Phi=\frac{1}{2}\left[G^2-(F+\mathcal{C})^2\right],
\end{equation}
where $\mathcal{C}\equiv\mathcal{C}(G,H,U)$ is an arbitrary function. Then,
\begin{equation}
\Phi_F=-(F+\mathcal{C}),\quad
\Phi_G=G-(F+\mathcal{C})\,\mathcal{C}_G,\quad
\Phi_H=-(F+\mathcal{C})\,\mathcal{C}_H,\quad
\Phi_U=-(F+\mathcal{C})\,\mathcal{C}_U.
\end{equation}
The trivial choice $\mathcal{C}=0$ makes $g=\theta$, $h=\nu$, and $u=t-\mu\,(\psi-e\sin\psi)/\sqrt{8U^3}$.
\par

Choosing $\mathcal{C}$ as a linear combination of the momenta, simply ads $\phi$ to the angle variables. Specifically, when we set $\mathcal{C}=-G$ we obtain\footnote{Note that Hill proceeds by  scaling $r$ by $1/\sqrt{G-\frac{1}{2}F}$, or choosing $\chi=r^2/(G-\frac{1}{2}F)$ in our notation. This is equivalent to scaling our $\Phi$ new Hamiltonian by $G-\frac{1}{2}F$, what yields to Hill's Hamiltonian $\Phi=F$ while the transformation (\ref{TRmapping}) remains unaltered.}
\begin{equation}\label{HillTR}
\Phi=F\left(G-\frac{1}{2}F\right)
\end{equation}
and recover the TR-mapping \cite{Scheifele,Deprit1981a}
\begin{equation}\label{TRmapping}
f=\phi,\qquad g=\theta-f,\qquad h=\nu,\qquad u=t-\frac{\mu}{\sqrt{8U^3}}\,(f-e\sin{f}).
\end{equation}

\section{Conclusions}

A deep insight in the computation of canonical transformations is obtained through a general formulation of the Hamilton-Jacobi equation in the extended phase space that, besides, includes a regularizing function. As far as the transformed Hamiltonian may remain formal, one can obtain families of canonical transformations instead of single sets of canonical variables. Then, particular transformations that meet the user's required characteristics are systematically computed from the solution of partial differential equations that may enjoy very simple solutions.
\par

Our formulation shows that different transformations in the literature belong to the same families. Furthermore, it reveals fundamental features of the transformations. Thus, among the worked examples, we identify families having as members known canonical changes of variables that originally were defined through implicit relations. 

\section*{Acknowledgements}
Partial support is recognized from projects MTM 2006-06961 (S.F.) and AYA 2009-11896 (M.L.) of the Government of Spain, and a grant from Fundaci\'on S\'eneca of the autonomous region of Murcia.


\end{document}